\documentclass[%
 reprint,
superscriptaddress,
 amsmath,amssymb,
 aps,
 showkeys,
prc,
floatfix,
]{revtex4-2}
\usepackage{dcolumn}
\usepackage{bm}
\usepackage{amssymb,amsmath,bm,mathtools}
\usepackage{siunitx,booktabs}
\usepackage{graphics,graphicx,epstopdf,epsfig}
\usepackage{tabularx}
\usepackage{psfrag}
\usepackage{diagbox}
\usepackage{xcolor}
\usepackage{array}
\usepackage[english]{babel}
\usepackage{hhline}
\usepackage{adjustbox}
\usepackage{nicematrix}
\usepackage{multirow}
\def\nbR{\ensuremath{\mathrm{I\! R}}}

\newcommand{\be}{\begin{equation}}
\newcommand{\ee}{\end{equation}}
\newcommand{\ba}{\begin{eqnarray}}
\newcommand{\ea}{\end{eqnarray}}
\begin{document}
\title{Determination of the moments of the proton charge density}

\author{M.~Atoui}
\affiliation{Universit\'e Paris-Saclay, CNRS/IN2P3, IJCLab, 91405 Orsay, France}

\author{M.B.~Barbaro}
\affiliation{Dipartimento di Fisica, Universit\'a di Torino and INFN Sezione di Torino, 10125 Torino, Italy}
\affiliation{IPSA-DRII, 94200 Ivry-sur-Seine, France}

\author{M.~Hoballah}
\affiliation{Universit\'e Paris-Saclay, CNRS/IN2P3, IJCLab, 91405 Orsay, France}

\author{C.~Keyrouz}
\affiliation{Universit\'e Paris-Saclay, CNRS/IN2P3, IJCLab, 91405 Orsay, France}
\affiliation{Universit\'e C\^ote d'Azur, Institut de Physique de Nice, 06200 Nice, France}

\author{M.~Lassaut}
\affiliation{Universit\'e Paris-Saclay, CNRS/IN2P3, IJCLab, 91405 Orsay, France}

\author{D.~Marchand}
\affiliation{Universit\'e Paris-Saclay, CNRS/IN2P3, IJCLab, 91405 Orsay, France}

\author{G.~Qu\'em\'ener}
\affiliation{Normandie Univ, ENSICAEN, UNICAEN, CNRS/IN2P3, LPC Caen, 14000 Caen, France}

\author{E.~Voutier}
\affiliation{Universit\'e Paris-Saclay, CNRS/IN2P3, IJCLab, 91405 Orsay, France}

\date{\today}

\begin{abstract}
A global analysis of proton electric form factor experimental data from Rosenbluth separation and low squared four-momentum transfer experiments is discussed for the evaluation of the spatial moments of the proton charge density based on the recently published integral method~\cite{Hob20}. Specific attention is paid to the evaluation of the systematic errors of the method, particularly the sensitivity to the choice of the mathematical expression of the form factor fitting function. Within this comprehensive analysis of proton electric form factor data, the moments of the proton charge density are determined for integer order moments, particularly: $\langle r^2 \rangle$=0.6{94(09)}$_{Sta.}${(16)}$_{Sys.}$~fm$^2$, $\langle r^3 \rangle$=0.{870(46)}$_{Sta.}${(80)}$_{Sys.}$~fm$^3$, and $\langle r^4 \rangle$=1.{47(25)}$_{Sta.}${(45)}$_{Sys.}$~fm$^4$.
\end{abstract}

%
%

\maketitle


%
%

\section{Introduction}

Two experimental techniques have been proposed and developed over decades of scientific research to determine the charge radius of the proton $R_p$ (see Refs.~\cite{Kar20,Gao22} for recent reviews): the spectroscopy technique where $R_p$ is determined from the hyperfine structure of hydrogen atoms, and the scattering technique where $R_p$ is deduced from the cross section of elastic lepton scattering off a proton target. By definition, $R_p$ is obtained from the slope of the proton electric form factor $G_E(k^2)$ at zero squared four-momentum transfer $k^2$  
\begin{equation}
R_p \equiv \sqrt{ -6 \left. \frac{\mathrm{d} G_E(k^2)}{\mathrm{d}k^2} \right\vert_{k^2=0} } \,
\label{redef} 
\end{equation}
which on the one hand is directly accessed from the energy levels of hydrogen atoms, and on the other hand is indirectly obtained from the measurements of $G_E(k^2)$~\cite{Car15,Mil19,Lor20}. Since lepton scattering cannot reach the zero squared four-momentum transfer limit, the scattering technique relies on the zero extrapolation of $G_E(k^2)$ and then strongly depends on the functional form as well as on the data analysis method used for the extrapolation. Consequently, the scattering technique is  believed to be intrinsically less accurate than the spectroscopy one. Indeed, the most precise measurements from muonic hydrogen spectroscopy 0.84184(67)~fm~\cite{Poh10} and 0.84087(39)~\cite{Ant13} are more than ten times more accurate than the best scattering measurements 0.879(8)~fm~\cite{Ber10,Ber14-1} and 0.831(14)~fm~\cite{Xio19}. This appears as a blatant limitation of the scattering technique, particularly difficult to overcome as proved by the many lepton scattering projects that developed following the advent of the so-called proton radius puzzle~\cite{Ber14}. It also hampers any attempt to determine higher-order moments of the proton charge density through this approach, here-after referred to as the derivative method.

{The release of new hydrogen spectroscopy measurements, two within a 1$\sigma$ agreement~\cite{Bey17,Bez19} with muonic hydrogen and one in 3.6$\sigma$ disagreement~\cite{Fle18}, motivated the 2018 CODATA evaluation of the proton radius to include muonic atom data~\cite{Tie21}. The 2018 CODATA so recommended the proton radius 0.8414(19)~fm which was further confirmed within 1.6$\sigma$ by a high precision measurement of the 1S-3S transition~\cite{Gri20}. However the latest result about the 2S-8D transition, within a 3.1$\sigma$ disagreement~\cite{Bra21} with CODATA, suggests that the issue is not yet settled.}

Addressing the determination of the proton charge radius from electron scattering data, we have recently proposed a novel approach~\cite{Hob20}, referred to in the following as the integral method, that enables the determination of spatial moments of any real-valued order $\lambda >-3$ through integral forms of the Fourier transform of the probability density function. 
\begin{table*}[t!]
\centering
\begin{tabular}{c|ccc|ccc} 
\hline\hline
Data &  &  &  & Number &   \multicolumn{2}{c}{$k^2$-range} \\
 Set & {Year} & {Authors} & {Ref.} & of & $k^2_{min}$  & $k^2_{max}$ \\
Number & & & & data points & [fm$^{-2}$]  & [fm$^{-2}$] \\ \hline\hline
 \phantom{1}1 & 1962 &         Lehmann {\it et al.} &   \cite{Leh62} & \phantom{1}1 & 2.98  & 2.98 \\
 \phantom{1}2 & 1963 &       Dudelzak {\it et al.} &   \cite{Dud63} & \phantom{1}4 & 0.30   & 2.00 \\
 \phantom{1}3 & 1963 &       Berkelman {\it et al.} &   \cite{Ber63} & \phantom{1}3 &  25.0 & 35.0 \\
 \phantom{1}4 & 1966 &   Fr\`erejacque {\it et al.} &   \cite{Fre66} & \phantom{1}4 & 0.98  & 1.76 \\
 \phantom{1}5 & 1966 &            Chen {\it et al.} &   \cite{Che66} & \phantom{1}2 & 30.0   & 45.0 \\
 \phantom{1}6 & 1966 &        Janssens {\it et al.} & \cite{Jan66} &           20 & 4.00    & 22.0 \\
 \phantom{1}7 & 1971 &          Berger {\it et al.} &   \cite{Ber71} &  \phantom{1}9           & 1.00 & 50.0  \\
8 & 1973 &         Bartel {\it et al.} &   \cite{Bar73} &   \phantom{1}8          & 17.2 &77.0 \\
9 & 1975 &       Borkowski {\it et al.} &   \cite{Bor75} &           10 & 0.35 & 3.15 \\
10 & 1994 &          Walker {\it et al.} &   \cite{Wal94} & \phantom{1}4 & 25.7 & 77.0 \\
11 & 1994 &       Andivahis {\it et al.} &   \cite{And94} & \phantom{1}8 & 44.9 & 226. \\
12 & 2004 &         Christy {\it et al.} &   \cite{Chr04} & \phantom{1}7 & 16.7 & 133. \\
13 & 2005 &           Qattan {\it et al.} &   \cite{Qat05} & \phantom{1}3 & 67.8 & 105. \\
14 & 2014 &        Bernauer {\it et al.} & \cite{Ber14-1} &   77 & 0.39 & 14.2 \\
15 & 2019 &  Xiong {\it et al.} - 1.1 GeV & \cite{Xio19} &     33      & 5.51$\times$10$^{-3}$ & 3.96$\times$10$^{-1}$  \\ 
16 & 2019 &    Xiong {\it et al.} - 2.1 GeV & \cite{Xio19} &       38     & 1.79$\times$10$^{-2}$  & 1.49 \\ 
17 & 2021 & Mihovilovi\v{c} {\it et al.} - 195 MeV &   \cite{Mih21} & \phantom{1}6 & 3.43$\times$10$^{-2}$ &  6.99$\times$10$^{-2}$ \\
18 & 2021 & Mihovilovi\v{c} {\it et al.} - 330 MeV &   \cite{Mih21} &           11 & 4.69$\times$10$^{-2}$ & 2.00$\times$10$^{-1}$\\
19 & 2021 & Mihovilovi\v{c} {\it et al.} - 495 MeV &   \cite{Mih21} & \phantom{1}8 & 1.57$\times$10$^{-1}$ & 4.37$\times$10$^{-1}$ \\
\hline\hline
\end{tabular}
\caption{Proton electric form factor data considered in the present study, as explained in the text.}
\label{ExpAll}
\end{table*}
Two techniques were proposed to regularize the Fourier integrals. The weak limit one is used here for its ability to deliver analytical expressions of moments of integer order. In this approach, the $\lambda$-order moment of the proton charge spatial density $\rho_E({\bf r})$ is defined as 
\be
( r^{\lambda}, \rho_E ) = \langle r^{\lambda} \rangle = \int_{\nbR^3}  \mathrm{d}^3\bm{r} \, r^{\lambda} \rho_E({\bm r}) \, 
\ee
and can be expressed as
\be
\langle r^{\lambda} \rangle = \frac{2}{\pi}  \, \Gamma(\lambda+2) 
\lim_{\xi \to 0^+} \int_{0}^{\infty} \mathrm{d}k \, {\mathcal I}_{\lambda}(k,\xi) \, G_E(k^2) \label{eqIM}
\ee
with
\be
{\mathcal I}_{\lambda}(k,\xi) = \frac{k \, \sin\left[{(\lambda+2) \, {\rm Arctan} \left( k/\xi \right)} \right] }{(k^2 + \xi^2)^{\lambda/2+1}}
  \, 
\ee
and where the form factor and the density are related by the Fourier transform 
\be
G_E(k^2) = \int_{\nbR^3} \mathrm{d}^3\bm{r} \, e^{- i \bm{k}\cdot \bm{r}} \rho_E({\bm r}) \, .  \label{FTlink}
\ee
The integral method overcomes the restriction of the derivative method which is limited to moments of positive even orders. Each moment order of the charge density is of interest as it carries complementary information on the charge distribution inside the proton. For instance, the short-distance behaviour of the charge distribution is encoded in the negative order moments which are particularly sensitive to the large $k^2$-dependence of $G_E(k^2)$, while the long-distance behaviour is encoded in the high positive order moments sensitive to the low $k^2$-dependence.

The integral method was demonstrated strictly equivalent to the derivative method for positive even order moments $\langle r^{2p} \rangle$~\cite{Hob20}, even though basic concepts distinguish them. While the derivative method crucially depends on the way the form factor approaches its value at zero $k^2$, the integral method involves the full $k^2$ physics region as expressed by the integral in Eq.~\eqref{eqIM}. This conceptual change is of importance for the determination of the functional form of $G_E(k^2)$ which can be expected more precise, and similarly for the moments. In practice the entire $k^2$ physics region may be restricted as the high $k^2$-region might not contribute significantly to the integral of Eq.~\eqref{eqIM}. This is particularly true for positive order moments which, considering the $k^2$-dependence of experimental data, are dominated by the region $k^2 \le 2$~GeV$^2$ (51.4~fm$^{-2}$). 

Hence, all available data should be considered for analysis when attempting evaluations of $R_p$ and more generally $\langle r^{\lambda} \rangle$. Following the formal demonstration of the integral method, the present study aims at the evaluation of a selected set of $\lambda$-order moments of the proton charge density from proton electric form factor data, namely odd and even orders in the range $-2 \le \lambda \le 7$. The next section describes the selection and the analysis of experimental data. The determination of the charge density moments and of their statistical significance is discussed in a further section. A specific attention is then paid to the evaluation of the systematic errors, and is followed by a dedicated discussion on the determination of the proton charge radius in the context of the integral method. 

%
%

\section{Analysis of $G_E(k^2)$ experimental data}
\label{section2}

The data inputs of the present work consist of pro-
ton electric form factor data $G_E(k^2)$ extracted from elec-
tron scattering experiments through a Rosenbluth sepa-
ration~\cite{Ros50} or for kinematical conditions where the contribution of the magnetic form factor ($G_M(k^2)$) to the cross
section is strongly suppressed, for instance at very low $k^2$.
For the purposes of the present work aiming at a quantitative evaluation of the merits of the integral method, experiments measuring polarized lepton beam observables
are not considered. Consequently, $G_E(k^2)$ is overestimated at large $k^2$, that is in a region where it already
has a small magnitude. The list of experiments selected
according to the previous criteria is reported in Table~\ref{ExpAll}
by chronological order. While modern experiments after
the nineties are easily classified with respect to these criteria, the selection appears more intricate for early elastic scattering experiments where data analysis sometimes
involves the empirical scaling $G_M(k^2)$=$\mu G_E(k^2)$ of the
proton electromagnetic form factors, or where the Rosenbluth separation combines different experiments. 
Additionally, some analyses report negative or null squared
form factors indicating some experimental issues, mostly
related to a small $\epsilon$ range in the Rosenbluth separation
where $\epsilon$ is the so-called polarization of the virtual photon.
The reduction of the published data set following the previous criteria concerns Refs.~\cite{Leh62,Che66,Jan66}, and specifically
Ref.~\cite{Ber63} for the constraint $\epsilon$$>$$0.1$. 
Earlier experimental
data of Refs.~\cite{Lit61,Bum61} were initially considered but finally
rejected because of their abnormal $k^2$-dependence: analyzing a data set including these experiments only degrades the fit quality while not significantly changing the
fit parameters. The relevant number of data points referring to a single experiment and passing the different
constraints mentioned above are indicated in the fifth column of Table~\ref{ExpAll}. The main differences between the data
sets, notably the investigated $k^2$-range, are reported in
Table~\ref{ExpAll}. Note that the PRad (Proton Radius)~\cite{Xio19} and ISR
(Initial State Radiation)~\cite{Mih21} experimental data are separated in 2 (PRad) and 3 (ISR) distinct sets corresponding
to different beam energies and squared four-momentum
transfer domains where the $G_M(k^2)$ contribution to the
cross section is negligible.

\begin{table}[t!]
\label{KelPara}
\begin{center}
\begin{tabular}{c|c|c|c}
\hline\hline
 $a_1$ & $b_1$ & $b_2$ & $b_3$ \\
 $\left[\times10^{-1}\right.$~fm$\left.^2\right]$ & $\left[\times10^{-1}\right.$~fm$\left.^2\right]$  & $\left[\times10^{-1}\right.$~fm$\left.^4\right]$ & $\left[\times10^{-3}\right.$~fm$\left.^6\right]$  \\ \hline
 8.80{08} & 9.9{570} & 1.0{285} & 2.{9252} \\
\hline\hline
$(\delta a_1)_{Sta.}$ & $(\delta b_1)_{Sta.}$ & $(\delta b_2)_{Sta.}$ & $(\delta b_3)_{Sta.}$ \\ 
$\left[\times10^{-1}\right.$~fm$\left.^2\right]$ & $\left[\times10^{-1}\right.$~fm$\left.^2\right]$  & $\left[\times10^{-1}\right.$~fm$\left.^4\right]$ & $\left[\times10^{-3}\right.$~fm$\left.^6\right]$  \\ \hline
 0.00{54} & 0.0{113}  & 0.0{058} & 0.0{666} \\ \hline \hline
 $(\delta a_1)_{Sys.}$ & $(\delta b_1)_{Sys.}$ & $(\delta b_2)_{Sys.}$ & $(\delta b_3)_{Sys.}$ \\ 
 $\left[\times10^{-1}\right.$~fm$\left.^2\right]$ & $\left[\times10^{-1}\right.$~fm$\left.^2\right]$  & $\left[\times10^{-1}\right.$~fm$\left.^4\right]$ & $\left[\times10^{-3}\right.$~fm$\left.^6\right]$  \\ \hline
0.009{5} &  0.0019  & 0.00{03} & 0.0{219} \\ \hline\hline
\end{tabular}
\end{center} 
\vspace*{-10pt}
\caption{Fit parameters of the $G_E(k^2)$ functional of Eq.~\eqref{kellymodel} with their associated statistical and systematic errors.}
\end{table}

In total, the experimental data set considered in the present study consists of 19 single electric form factor data sets up to $k^2$=226~fm$^{-2}$ (8.8~GeV$^2$). The complete set is analyzed within a simultaneous fit approach requiring the same $k^2$-dependence for each of the experiments and a separate normalization parameter factor for each single data set. This is expressed in the polynomial ratio~\cite{Kel04}
\begin{equation}
\label{kellymodel}
{G_E^i(k^2) \equiv \eta_i \, G_E(k^2)} = \eta_i\, \frac{1+a_1 k^2}{1+b_1 k^2 + b_2 k^4 + b_3 k^6}
\end{equation}
where $\eta_i$ is the normalization fit parameter of the data set number $i$. Note that the use of a $k^2$-constant normalization parameter follows the analysis techniques of the most recent experiments~\cite{Ber14-1,Xio19,Mih21} and remains appropriate even for high $k^2$ data. This $k^2$-region, which features form factors of very small magnitude, is of importance for the determination of negative order moments where systematic effects of the integral {method} dominate~\cite{Hob20}. The results of the best fit to experimental data are reported in Tab.~\ref{KelPara} and Tab.~\ref{tab_normalisationall} in terms of the $(a_1,b_1,b_2,b_3)$ parameters of the form factor function and of the $\eta_i$ normalization fit parameters, respectively. Statistical errors are determined taking into account the correlations between each parameter. Systematic errors of the fit parameters reflect experimental data systematics.

The systematics are determined by shifting upwards or downwards the data points of each data set with their respective systematics. A total of {2$^{19}$} configurations were considered, corresponding to each possible data set combination. For each configuration, the data points of a given data set are all shifted in the same direction while each data set is independently shifted upwards or downwards. The systematics on each parameter is obtained from the arithmetic average of the distribution of the difference between the parameter values of the fit of the shifted data and the ones of the reference fit in the first line of Tab.~\ref{KelPara}. 

\begin{table}[t!]
\begin{tabular}{c|c|ccc}
\hline\hline
 Data Set & \multirow{2}{*}{Ref.} & \multirow{2}{*}{$\eta_i$} & {$(\delta \eta_i)_{Sta.}$}  & {$(\delta \eta_i)_{Sys.}$} \\
Number & & & {$\left( \times 10^{-2} \right)$} & {$\left( \times 10^{-2} \right)$} \\ \hline
 \phantom{1}1 & \cite{Leh62}   & 0.993 & {3.020} & 9.92{2} \\
 \phantom{1}2 & \cite{Dud63}   & 0.98{2} & 0.{505} & 0.752 \\
 \phantom{1}3 & \cite{Ber63}   & 2.{441} & {15.87} & 1{2.20} \\
 \phantom{1}4 & \cite{Fre66}   & 0.991 & 0.{917} & 0.208 \\
 \phantom{1}5 & \cite{Che66}   & 0.{922} & {30.43} & 4.{612} \\
 \phantom{1}6 & \cite{Jan66}   & 1.00{4} & 1.1{32} & 0.80{3} \\
 \phantom{1}7 & \cite{Ber71}   & 1.00{1} & {1.333} & {2.001} \\
 \phantom{1}8 & \cite{Bar73}   & 1.0{25} & {4.490} & 1.0{77} \\
 \phantom{1}9 & \cite{Bor75}   & 0.981 & 0.{254} & 1.766 \\
           10 & \cite{Wal94}   & 1.1{70} & {4.902} & 0.{469} \\
           11 & \cite{And94}   & 0.9{72} & {2.144} & 6.{812} \\
           12 & \cite{Chr04}   & 1.0{42} & {3.513} & 0.5{06} \\
           13 & \cite{Qat05}   & 1.0{72} & {3.509} & 0.{584} \\
           14 & \cite{Ber14-1} & 0.991 & 0.0{83} & 0.99{3} \\
           15 & \cite{Xio19}   & 1.000 & 0.0{22} & 0.215 \\ 
           16 & \cite{Xio19}   & 0.998 & 0.0{18} & 0.119 \\ 
           17 & \cite{Mih21}   & 1.001 & 0.{113} & 0.370 \\
           18 & \cite{Mih21}   & 1.000 & 0.0{97} & 0.365 \\
           19 & \cite{Mih21}   & 0.998 & 0.0{66} & 0.4{42} \\
\hline\hline
\end{tabular}
\caption{Normalization fit parameters corresponding to each data set considered in this study, together with their statistical and systematic errors determined as explained in the text.}
\label{tab_normalisationall}
\end{table}

\begin{figure}[t!]
\includegraphics[width=\columnwidth]{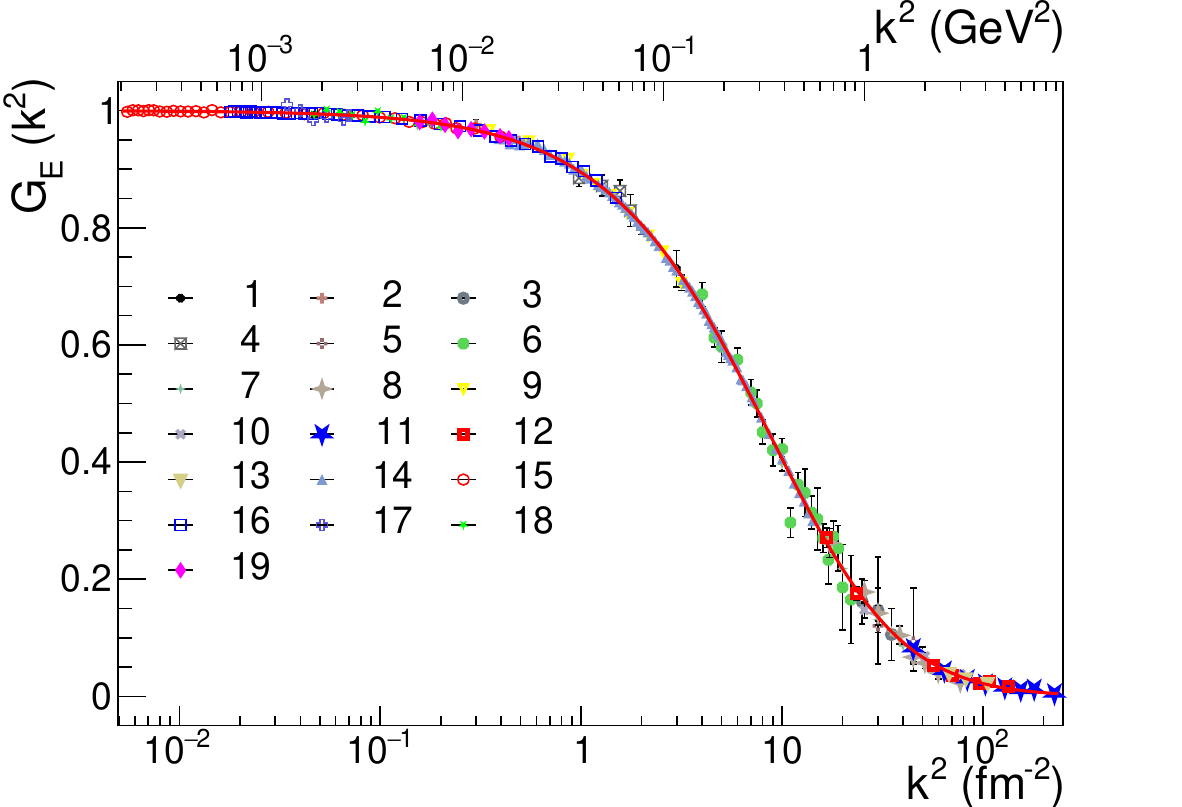}
\caption{Simultaneous fit (black line) of {normalized experimental data ($G_E^i(k^2)/\eta_i$)} conside\-red in Tab.~\ref{ExpAll} using the polynomial ratio function of Eq.~\eqref{kellymodel}. Experimental data are labelled according to the numbering of the data set (see Tab.~\ref{ExpAll}).}
\label{FFit} 
\end{figure}

The deviation from 1 of the normalization fit parameters of most experiments is smal\-ler than 3\% and even 1\% for recent experiments~\cite{Ber14-1,Xio19,Mih21} which have the strongest statistical weight in the fitting procedure. The renormalization of raw PRad- and A1-Rosenbluth data, as obtained from the present simultaneous fit, corrects for their incompatibility at the high $k^2$-end of the PRad set. A few experiments~\cite{Che66,Wal94,Qat05} feature larger deviations up to 17\%, still reasonable at large $k^2$ where the contribution of the electric form factor to the cross section is small. Solely the experiment of Ref.~\cite{Ber63} requires abnormally large normalization fit parameter. No experimental peculiarities have been found in this work, which may explain such a large value, but the deviation of the corresponding data points from the global trend of other experiments is particularly striking in Fig.~\ref{FFit} which shows the best fit (Tab.~\ref{KelPara}) to experimental data.

\begin{figure}[t!]
\includegraphics[width=\columnwidth]{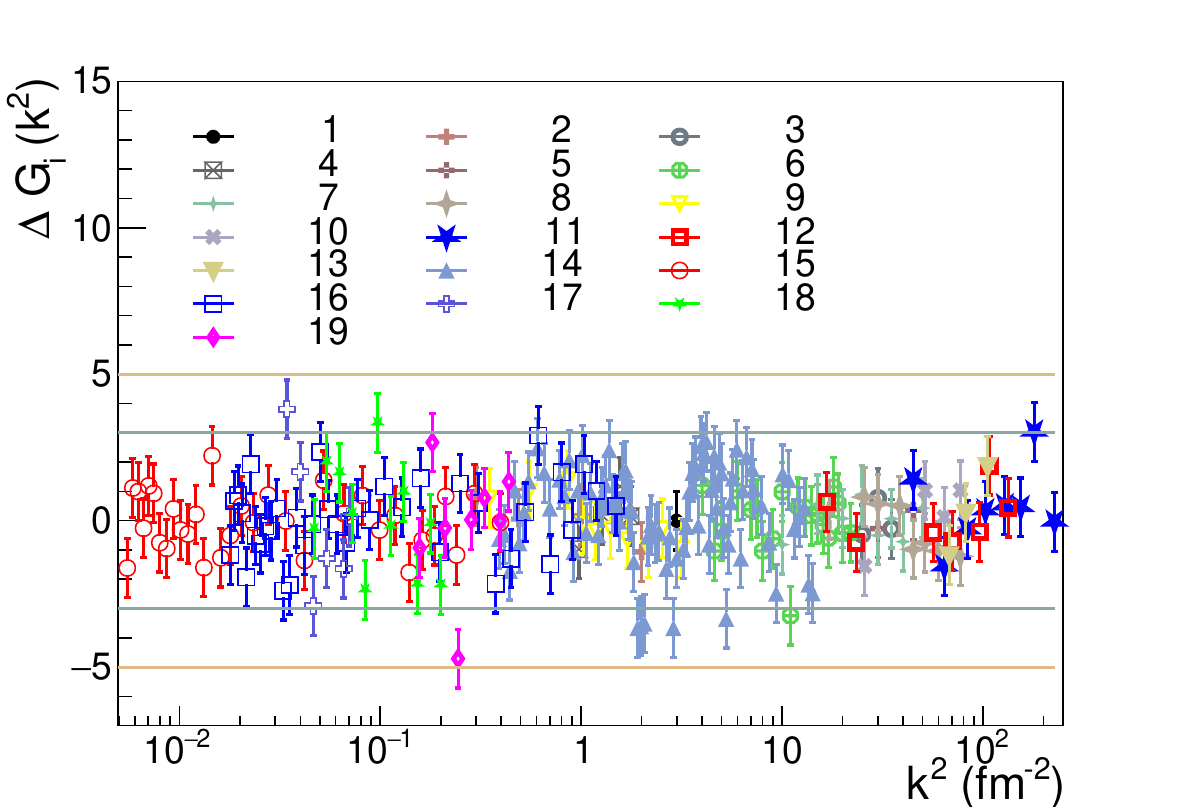}
\caption{Residual deviations of experimental data with respect to the fit predicted value considering their corresponding normalization fit parameter, as defined in Eq.~\eqref{eq:res} in units of the statistical standard deviation of each data point. The $\pm$3 and $\pm$5 standard deviation limits are indicated by horizontal lines. Data labels are identical to Fig.~\ref{FFit}.}
\label{Fres} 
\end{figure}

This best fit accounts for a reduced $\chi^2_r$ {of 1.97}, which is reasonable considering the actual dispersion of data. The residual deviation $\Delta G_i(k^2)$ of experimental data from the data set $i$ can be defined as 
\begin{equation}
\Delta G_i(k^2) = \frac{G_E^i(k^2) - \eta_i G_{{E}}(k^2)}{\delta G_E^i(k^2)} 
\label{eq:res}
\end{equation}
where $G_E^i(k^2)$ is the experimental data of the set $i$ with its corresponding statistical error $\delta G_E^i(k^2)$ and normalization fit parameter $\eta_i$, and $G_{{E}}(k^2)$ is the fit predicted value. The distribution of $\Delta G_i(k^2)$ (Fig.~\ref{Fres}) appears consistent with the obtained $\chi^2_r$ value, {that is in the acceptable $\pm$3$\delta G_E^i(k^2)$ range.} 
 
%
%

\section{Determination of experimental moments}
\label{section3}

The spatial moments of the proton charge density are determined from the form factor function using the integral method restricted to the measured physics region. This is expressed in terms of truncated moments, where the integral region of Eq.~\eqref{eqIM} is limited to a finite upper boundary $Q$, that is
\begin{equation}
\langle r^{\lambda} \rangle_{_Q} = \frac{2}{\pi}  \, \Gamma(\lambda+2) \, 
\lim_{\xi \to 0^+} \int_{0}^{Q} \mathrm{d}k \, {\mathcal I}_{\lambda}(k,\xi) \, G_E(k^2) \, . \label{eqIMT}
\end{equation}
The full or infinite moment of order $\lambda$ can thus be formally expressed as
\be
\langle r^{\lambda} \rangle = \lim_{Q \to \infty} \langle r^{\lambda} \rangle_{_Q} \, . \label{eq:mominf}
\ee
In the absence of experimental data at very large $k^2$, pertubative Quantum ChromoDynamics provides scaling rules which predict a rapid decrease of $G_E(k^2)$~\cite{Bro73}. Thus, the effect of the truncation of the integral in Eq.~\eqref{eqIMT} can be controlled. The integration cut-off $Q$=7.2~fm$^{-1}$ (corresponding to $Q^2$=51.4~fm$^{-2}$=2 GeV$^{2}$) considered in the  present study was proven to only impact the evaluation of negative order moments~\cite{Hob20}. Assuming a physically acceptable  functional form, {\it i.e.} without any pole in the physics region, the considered form factor parameterization can be conveniently decomposed as a sum of complex monopoles according to the expression   
\ba
G_E(k^2) &=& \frac{1}{b_3} \, \frac{1+a_1k^2}{(k^2-k_1^2)(k^2-k_2^2)(k^2-k_3^2)} \nonumber \\
&=& \sum_{n=1}^3 \frac{A_n}{k^2+(ik_n)^2} \label{ffefun}
\ea
with 
\be
A_n = - \frac{1}{b_3} \, \frac{1 - a_1 k_n^2}{\prod_{m=1,m \neq n}^3 (k_n^2 - k_m^2) } \, 
\ee
and where the $k_n$'s are the complex roots of the equation
\be
(k^2)^3 + \frac{b_2}{b_3} \, (k^2)^2 + \frac{b_1}{b_3} \, k^2 + \frac{1}{b_3} = 0 \, .
\ee
Correspondingly, the truncated moments can be written as

\be
\left( r^{\lambda},\rho_{R_3^1} \right)_{_Q} = - \sum_{n=1}^3 \frac{A_n}{k_n^2} \, \left( r^{\lambda},\rho_{M \left( i k_n \right)} \right)_{_Q}
\ee
where the density index $R^1_3$ denotes the form factor parameterization with which it is Fourier related (Eq.~\eqref{FTlink}), {\it i.e.} $R^1_3$ stands for the ratio of a polynomial of order 1 with a polynomial of order 3 in $k^2$, and $M(ik_n)$ for a monopole with parameter $ik_n$ ($\rho_{M \left( i k_n \right)}=1/(k-ik_n)$). The odd and even truncated monopole moments write 

\begin{table*}[t!]
\begin{center}
\begin{tabular}{c||c|c|c||c|c||c|c|c|c|c}
\hline\hline
 & &  &  & \multicolumn{2}{c||}{Statistical Error} & \multicolumn{5}{c}{Systematic Error} \\
 {$\lambda$} & {$\langle r^{\lambda} \rangle_{_Q}$} & {$\langle r^{\lambda} \rangle$}  &{${\langle r^{2p} \rangle}_{d}$} & $\delta \left[ {\langle r^{\lambda} \rangle_{_Q}} \right]$ & $\delta \left[ {\langle r^{2p} \rangle_{d}} \right]$ & Dat. & Int. & Fun. & Nor. & Mod.\\  
 & $\left[\mathrm{fm}^{\lambda}\right]$ & $\left[\mathrm{fm}^{\lambda}\right]$ & $\left[\mathrm{fm}^{\lambda}\right]$ & $\left[\mathrm{fm}^{\lambda}\right]$ & $\left[\mathrm{fm}^{\lambda}\right]$ & $\left[\mathrm{fm}^{\lambda}\right]$ & $\left[\mathrm{fm}^{\lambda}\right]$ & $\left[\mathrm{fm}^{\lambda}\right]$ & $\left[\mathrm{fm}^{\lambda}\right]$ & $\left[\mathrm{fm}^{\lambda}\right]$\\ \hline
          -2 & 6.5{245}   & 8.{7226}   &               $-$ & 0.0{172}   & $-$    & 0.01{06}  & 2.{1981} &  0.000{1}  &   {0.0095}   & 0.{3731}   \\
          -1 & 1.9{681}   & 2.{0906}   &               $-$ & 0.00{24}   & $-$    & 0.00{19}  & 0.1{225} &  0.000{1}  &   {0.0029} & 0.0{278}    \\
\phantom{-}1 & 0.7{206}   & 0.71{79}   &               $-$ & 0.00{20}   & $-$    & 0.0025  & 0.00{27}            &  0.0001  &   {0.0011}  & 0.00{29} \\
\phantom{-}2 & 0.6{937}   & 0.6{937}   & 0.6{937} & 0.00{94}   & 0.0{105} & 0.011{1}  & $0$               &  0.0001  &   {0.0010} & 0.0{116} \\
\phantom{-}3 & 0.{8697}   & 0.{8701}   &               $-$ & 0.0{457}   & $-$    & 0.0{494}  & 0.0004 &  0.000{7}  &   {0.0013}  & 0.0{633}    \\
\phantom{-}4 & 1.{4728}   & 1.{4728}   & 1.{4728} & 0.{2461}   & 0.{2365} & 0.24{74}  & $0$               &  0.00{65}  &   {0.0022}  & 0.{3805}  \\
\phantom{-}5 & {3.8139}   & {3.8137} &               $-$ & {1.4822}   & $-$    & 1.4{297}  & 0.0002            &  0.0{343}  &   {0.0056}  & {2.6276}    \\
\phantom{-}6 & {16.405}  &              {16.405}  &           {16.405} & {10.058} & {10.839} & 9.{4985}  & $0$               &  0.{1871}  &   {0.0240}  & {21.531} \\
\phantom{-}7 &             {104.64} &              {104.64} &             $-$ & {76.676}  & $-$    & 71.{727} & 0.000{1}            &  {2.8169}  &   {0.1528} & {212.48}    \\
\hline\hline
\end{tabular}
\end{center}
\caption{Moments of the proton charge density as determined from the integral method and the derivative one for even moments. The truncated moments of the second column are evaluated for the cut-off $Q^2$=52~fm$^{-2}$ and the infinite moments of the third column are similarly evaluated in the limit $Q \to \infty$, assuming that the large $k^2$-dependence of the form factor is described by Eq.~\eqref{kellymodel}. Even moments obtained by the derivative method are presented in the fourth column with their associated statistical errors in the sixth column. The different systematic error labels tag their origin: the data (Dat.), the integral method (Int.), the functional form (Fun.), {the data normalisation parameters (Nor.),} and the fitting model (Mod.) as described in the text.}
\label{Momres}
\end{table*}

\ba
& & \left( r^{-2},\rho_{M(ik_n)} \right)_{_Q} = \frac{(ik_n)^2}{2} \, \ln{\left( 1 + \frac{Q^2}{(ik_n)^2} \right)} \label{NegMom} \\
& & \left( r^{2p-1},\rho_{M(ik_n)} \right)_{_Q} = \frac{2}{\pi} \, \frac{(2p)!}{(ik_n)^{2p-1}} \label{OddMom} \\
& & \,\,\, \times \,  \left\{ {\mathrm{Arctan}{\left( \frac{Q}{ik_n} \right)} } - \sum_{j=1}^{p} \frac{(-1)^j}{2j-1} {\left( \frac{ik_n}{Q} \right)}^{2j-1} \right\} \nonumber \\
& & \left( r^{2p},\rho_{M(ik_n)} \right)_{_Q} = \frac{(2p+1)!}{(ik_n)^{2p}} \, . \label{2pmom}
\ea
The truncated moments corresponding to the form factor function of Eq.~\eqref{kellymodel} are reported in Tab.~\ref{Momres}, together with their asymptotic or so-called infinite value corresponding to the limit $Q \to \infty$. Note that the 0$^{\mathrm{th}}$-order moment is $1$ by definition of the fit function. As expected, negative moments are the only ones to significantly suffer from the truncation of the moment integral. Positive even moments are also compared to their values as obtained through the derivative method, determined from the expression 
\be \label{eq:der} 
{\langle r^{2p} \rangle}_{d} = (-1)^p \, \frac{(2p+1)!}{p!} \, \left. \frac{\mathrm{d}^p G_E(k^2)}{\mathrm{d} (k^2)^p} \right\vert_{k^2=0} \, . \ee
As demonstrated in Ref.~\cite{Hob20}, the derivative and integral determination of even moments must provide the same value, that is
\be
{\langle r^{2p} \rangle} = {\langle r^{2p} \rangle}_{d} \, .
\ee
This extends to truncated moments of any form factor function through the relation
\be
{\langle r^{2p} \rangle} = \langle r^{2p} \rangle_{_Q}
\ee
valid for any $Q\neq0$ (see Appendix~\ref{demo_trunc}). It is expressed in Eq.~(\eqref{2pmom}) and numerically verified in Tab.~\ref{Momres}.

The statistical errors of the experimental moments are determined from the propagation of the statistical errors of the fit parameters taking into account their correlations. The evaluation is based on the generation of 5$\times 10^4$ replica parameters  $P^i=(a_1^i,b_1^i,b_2^i,b_3^i)$ obtained from Gaussian distributions with mean values $\mu=(a_1,b_1,b_2,b_3)$ and covariance matrix $\Sigma$ as given by the fit of $G_E(k^2)$ experimental data. Each replica is built from the relationship 
\be
\label{eq:cov1}
P^i = C \cdot Z^i + \mu 
\ee
where $C$ is the triangular matrix such that
\be
 \Sigma = C \cdot C^T \,
\ee
with $C^T$ the transposed matrix, and  $Z^i=(\tilde{a}_1^i,\tilde{b}_1^i,\tilde{b}_2^i,\tilde{b}_3^i)$ is a random set of uncorrelated parameters obtained from Gaussian distributions with unit variances and zero means. The statistical errors reported in Tab.~\ref{Momres} correspond to the width of the distribution of replica moments. 

The magnitude of the statistical error limits the significance of the moments  determination to $\lambda<6$. This is a consequence of the existing data set which lacks measurements at ultra low $k^2$. High order positive moments indeed probe the long distance behaviour of the charge density and are therefore specifically sensitive to this region. High accuracy measurements in this region are extremely challenging. \newline
Concerning positive even moments, statistical errors of truncated moments can also be compared to those of the moments obtained from the derivative method. Considering the functional determined from the large data set selected in this study, the derivative method provides similar statistical errors as the integral method. As expected, the statistical uncertainty is obviously much better than the one that could be obtained from one single experiment. 

%
%

\section{Evaluation of systematic errors}
\label{section4}

The study developed hereafter intends to obtain as precise and realistic as possible the determination of systematic errors of the moments, a feature of upmost importance especially in the context of the proton radius determination. {Five} different sources of systematics are considered: the one related to the systematic error of the form factor measurements (Dat. in Tab.~\ref{Momres}), the one related to the determination method of the moments (Int. in Tab.~\ref{Momres}), the one intrinsic to the fit function (Fun. in Tab.~\ref{Momres}), {the one related to the data normalisation parameters of the fit function (Nor. in Tab.~\ref{Momres}),} and a last one attached to the choice of the fit function (Mod. in Tab.~\ref{Momres}).

The only experimental source of systematics corresponds to the error on the moment originating from the systematic errors of the measurements. It propagates to the moments through the systematics of the fit parameters and is determined by shifting upwards or downwards each parameter value with its systematic error. This leads to $2^4$ possible shifted parameter configurations for which the corresponding moments are determined and compared to the unshifted reference value. The error reported in the seventh column of Tab.~\ref{Momres} is the arithmetic average of the single difference evaluations. 

The other systematics are specifically attached to the integral method {\it i.e.} do not have an experimental origin {\it per se}. The first of them corresponds to the under- or over-estimation of the moments due to the truncation of the form factor integral. It is obtained from the comparison of the infinite (third column of Tab.~\ref{Momres}) and truncated moments (second column of Tab.~\ref{Momres}) and is reported in the eighth column of Tab.~\ref{Momres}. \newline
The second is the error intrinsic to the fit function, that is the bias generated on the fit parameters from the model itself. The error evaluation method consists in generating {form-factor} pseudo-data at the exact $k^2$ of experimental data but with a form factor value centered on the fit value $G_E(k^2)$. Each pseudo-data is then distributed according to a Gaussian with variance corresponding to the statistical error of real data. The resulting pseudo-data set is fitted with the function of Eq.~\eqref{kellymodel} to provide new parameters that are used to compute truncated moments. The procedure is repeated 5$\times 10^4$ times to generate the distributions of the moments from pseudo-data sets whose mean values are compared to the real data moments to yield the fit function systematics reported in the ninth column of Tab.~\ref{Momres}. \newline
{The third source of non-purely experimental systematic is attached to the normalization of experimental data (Tab.~\ref{tab_normalisationall}). It corresponds to the bias induced by the hypothesis that each data set may suffer from a global normalization issue as expressed by the $\eta_i$ parameters. The impact of this hypothesis on the determination of density moments is evaluated by comparing the $n$-order reference moments of the second column of Tab.~\ref{Momres} ($M^n_{ref.}$) with the average of the moments of each original data set ($\overline{M^n}$) defined as  
\begin{equation}
\overline{M^n} = M^n_{ref.} \, \frac{ \sum\limits_{i=1}^{19} \eta_i \, {\left[ (\delta \eta_i)_{Sta.}^2 + (\delta \eta_i)_{Sys.}^2 \right]}^{-1} }{\sum\limits_{i=1}^{19} { \left[ (\delta \eta_i)_{Sta.}^2 + (\delta \eta_i)_{Sys.}^2 \right]}^{-1}} \, .
\end{equation}
The difference $\vert \overline{M^n}-M^n_{ref.} \vert$ is reported in the tenth column of Tab.~\ref{Momres}.
}
\newline
The last source of systematics originates from the fitting model. The choice of the mathematical formulation of the fit function has indeed an impact on the determination of the moments~\cite{Yan18}. The evaluation of this systematics may however suffer from bias. One may try different fit functions and select the one {\it best} reproducing the experimental data, in which case the model systematics is zero independently of the bias in the definition of the {\it best} fit. Another approach is followed here where a {candidate} fit to experimental data is defined as any fit function featuring {$\chi^2_r \le 1.97$, that is a $\chi^2_r$ equivalent to that of the reference fit}. Several functional forms were investigated: polynomial of different orders, inverse polynomial of different orders, ratio of polynomials of different orders, and constant fraction expansion with different number of parameters. The only functions passing the selection criteria are the inverse polynomial of order 2 in $k^2$ and a constant fraction expansion with 3 {or 6} parameters. The error reported in the {eleventh} column of Tab.~\ref{Momres} is the difference between the average moments extracted from the functions passing the selection criteria and the reference moments (second column of Tab.~\ref{Momres}).

Except for negative moments where truncation effects are dominant, the model choice is a significant contribution to systematic errors of the integral method. It is worth noticing that this error is sometimes omitted, particularly in some of the many analyses of experimental data aimed at the determination of the charge radius of the proton after the highlighting of the proton radius puzzle~\cite{Ber14}. Using a mathematical function obtained from a physics model is the only way to minimize this error. 

%
%

\section{Determination of the proton charge radius}

As stressed in Sec.~\ref{section3} and experimentally verified in Tab.~\ref{Momres}, the integral method is strictly equivalent to the derivative method for even order moments. Particularly, we have for the second order moment 
\begin{equation}
\langle r^{2} \rangle = \langle r^{2} \rangle_{_Q} = -6 \left. \frac{\mathrm{d} G_E(k^2)}{\mathrm{d}k^2} \right\vert_{k^2=0}
\end{equation}
which, through the integral representation (Eq.~\eqref{eqIMT}) of the derivative, provides a novel way for the determination of the proton charge radius from experimental data. It is instructive to look at the evolution of the $R_p$ value over the years. Different groups of data sets are considered for this purpose, starting from a reference group gathering experimental data up to 1994 and then constituting new groups by successively adding new data sets. These groups are specifically defined in Tab.~\ref{TimeSet}. For each of them, the complete fitting procedure previously described is performed again, that is: determination of the parameters of the function of Eq.~\eqref{kellymodel}, evaluation of statistical and systematic errors of the parameters following the method described in Sec.~\ref{section2}, determination of the 2nd-order moment together with its statistical error and systematics according to Sec.~\ref{section3} and ~\ref{section4} respectively. The corresponding proton radius values and errors are reported in Tab.~\ref{TimeSet}. 

\begin{table}[t!]
\begin{center}
\begin{tabular}{c||c||c|c|c}
\hline\hline
  & Data set  &  $R_p$ & $(\delta R_p)_{Sta.}$ & $(\delta R_p)_{Sys.}$ \\  
 {\raisebox{5pt}{Time period}} &  range&  [fm] & [fm] & [fm] \\ [0.8ex]\hline
{1962}-1994 & 1 - {11} & 0.9{081} & 0.01{78} & 0.{1249} \\
{1962}-2005 & 1 - {13} & 0.{8813} & 0.01{91} & 0.{1044} \\
{1962}-2014 & 1 - {14} & 0.88{37} & 0.0{148} & 0.05{44} \\
{1962}-2019 & 1 - {16} & 0.8{329} & 0.00{57} & 0.0{102} \\
{1962}-2021 & 1 - {19} & 0.8{329} & 0.00{56} & 0.00{97} \\ \hline\hline
\end{tabular}
\end{center}
\caption{Evolution along the years of the proton charge radius value as a function of the group of data sets considered for each time period.}
\label{TimeSet}
\end{table}

The analysis of form factor data up to 2014 provides a proton charge radius in excellent agreement with the A1 Collaboration result~\cite{Ber10} and with the latest determination from the A1 cross section data only~\cite{Gra22}, however with much larger systematics. It is only taking into account the PRad~\cite{Xio19} measurements that the same analysis method yields a result within a 2$\sigma$ agreement with the measurements of muonic hydrogen~\cite{Poh10,Ant13}. It is worth noticing that all these evaluations of $R_p$, from early to the most recent experiments, are consistent with each other once systematic errors are taken into account. 

\begin{figure}[t!]
\vspace*{-17pt}
\begin{center}
\includegraphics[width=\columnwidth]{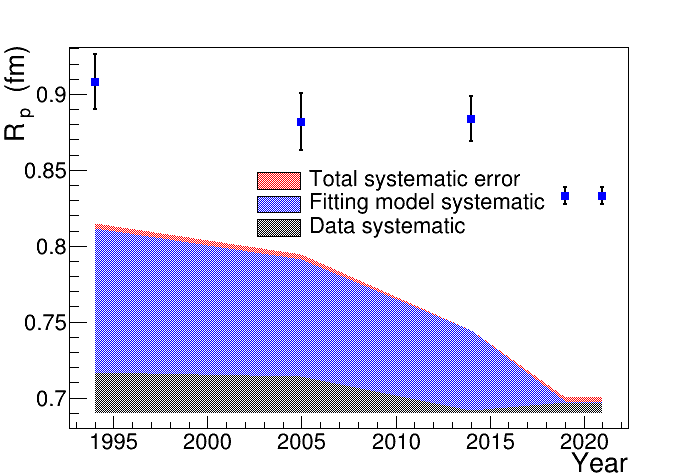}
\caption{Evolution of the proton charge radius value as determined with the integral method, considering the different groups of data sets defined in Tab.~\ref{TimeSet}. Each $R_p$ point is reported at the publication year of the last data set. Error bars are only statistical, and the magnitude of systematics is indicated by the shaded bands.}
\label{Rp} 
\end{center}
\end{figure}

The main sources of systematics are represented by shaded bands in Fig.~\ref{Rp}. The choice of the form factor function appears to be the essential contribution to $R_p$-systematic for the groups of measurements up to 2005. Its magnitude was strongly reduced by the availability of accurate low $k^2$ data in 2014. It is only with the latest PRad data that systematics become dominated by the component attached to experimental data. The present study of elastic electron scattering data allows the determination of $R_p$ with a 0.9\% accuracy, a remarkable number for electron scattering but still more than ten times less accurate than muonic hydrogen spectroscopy. The consistency within error bars of the different evaluations of $R_p$ further suggests that an underestimation of the systematics in the previous determinations of the proton charge radius is most likely responsible for the reported disagreement of electron scattering and muonic atom spectroscopy. We finally note that the present analysis yields the proton charge radius value 0.8{329}$\pm$0.00{56}$\pm$0.00{97}~fm in a {0.61}$\sigma$ agreement with the 2018 CODATA value~\cite{Tie21} from atomic spectroscopy {and} with the latest dispersion analysis of the electromagnetic form factors of the nucleon~\cite{Lin22}.

Whether $R_p$ could be interpreted as the {\it true} charge radius of the proton in a quantum sense is now a resolved issue. While it has been established that the spectroscopy and scattering techniques measure the same quantity in terms of the derivative of $G_E(k^2)$ (Eq.~\eqref{redef})~\cite{Mil19}, it is now recognized that this quantity does not have a strict probabilistic interpretation because of the relativistic nature of the proton. It was further argued that for proton-like objects, where the intrinsic size is comparable to the associated Compton wavelength, a charge density distribution cannot be unambiguously defined~\cite{Jaf21}. Several prescriptions have been suggested, like the infinite-momentum approach showing that the Dirac form factor $F_1(k^2)$ might be more appropriate than $G_E(k^2)$~\cite{Mil19}, however assimilating the proton to a disk. Another approach proposing a new definition of the electromagnetic spatial densities also prefers the use of $F_1(k^2)$, however quantitatively differing from the infinite-momentum approach~\cite{Pan22}. A recent work demonstrated that an unambiguous relativistic correction ($[1+k^2/4M^2]^{-1/2}$ with $M$ the nucleon mass) to the conventional Sachs distributions should be considered to justify their interpretation as rest-frame distributions, providing a natural interpolation between the Breit frame and the infinite-momentum frame distributions~\cite{Lor20,Che23}. This is consistent with earlier work advocating the same Darwin-Foldy correction $3/4M^2$ in the determination of nuclear sizes~\cite{Yen57,Fri97}. Based on this relativistic correction, we expressed in Appendix~\ref{momrel} the relativistic moments corresponding to the $G_E(k^2)$ function of Eq.~\eqref{ffefun}. We incidentally note that the {\it proton radius} so obtained as $\sqrt{\langle r^2 \rangle}$ is 0.8{526}$\pm$0.00{56}$\pm$0.00{97}~fm, where the first statistical error and the second systematic are determined following the methods described in the previous sections and using Eq.~\eqref{MomEvRel}.

%
%

\section{Conclusion}
\label{section5}

Based on a comprehensive analysis of the proton electric form factor data obtained from Rosenbluth separation and low $k^2$ measurements, the present work proposes an evaluation of the spatial moments of the proton charge density. A specific attention has been paid to the determination of statistical errors taking into account the correlations between the parameters of the form factor data fit. The actual status of experimental data allows a meaningful determination of the moments from $\lambda > -3$ up to the fifth order. 

Similarly, the designation of the different sources of systematic errors and their evaluation are thoroughly studied. Within the integral method evaluation of the moments, the sensitivity to the specific mathematical expression of the form factor fit function and the actual experiment systematics are found to be the most significant contributions to the systematic error of the moments. Taking into account the fit function sensitivity appears to reconciliate the different determinations of the proton charge radius. In that respect, the integral method approach yields the proton charge radius value 0.8{329}$\pm$0.00{56}$\pm$0.00{97}~fm. The current analysis suggests that the precision of this result would be improved by enriching the electric form factor data set at very low $k^2$ and using form factor functions supported by physics models.

%
%

\section*{Acknowledgement}

We thank J. van de Wiele for fruitful discussions. This work has received funding from the European Union’s Horizon 2020 research and innovation program under grant agreement No. 824093.

%
%

\appendix
\section{Even truncated moments} \label{demo_trunc}
In this Appendix, we will show that the even-order moments do not depend on the truncation domain. 
The moments of the density function $f$ at any integer order $n$$>$-$2$ can be expressed as~\cite{Hob20}
\ba
\label{ImAna}
( r^n, f ) & = & \frac{2}{\pi} \, \Gamma(n+2) \\ 
& \times & \lim_{\xi \to 0^+} \xi^{n+2} \int_0^{\infty} dk \, \tilde{f}(k) \, \frac{k \, \Phi_{n}(k,\xi)}{ (k^2 + \xi^2)^{n+2} } \nonumber 
\ea
with
\begin{equation}
\Phi_{n}(k,\xi) =\sum_{j=0}^{n+2} \sin \left( \frac{j \pi}{2} \right) \frac{(n+2)!}{j!(n+2-j)!} \, \left( \frac{k}{\xi} \right)^j \label{Phin} 
\end{equation}
and 
\begin{equation}
\tilde{f}(k) = \int_{\nbR^3} \mathrm{d}^3\bm{r} \, e^{- i \bm{k}\cdot \bm{r}} f({\bm r}) \, . 
\end{equation}
In terms of the binomial coefficients $C$, Eq.~\eqref{Phin} can be rewritten
\ba 
\Phi_{n}(k,\xi) & = & {\Im}{\mathrm{m}} \left[ \sum_{j=0}^{n+2} \exp \left( i \frac{j \pi}{2} \right)\, C_{n+2}^j \left( \frac{k}{\xi} \right)^j \ \right]  \\
& = & {\Im}{\mathrm{m}} \left[ \left(1 +  i \frac{k}{\xi}\right)^{n+2} \right] \, . \label{TheIM}
\ea
Reporting Eq.~\eqref{TheIM} into Eq.~\eqref{ImAna}, one obtains
\ba
( r^n, f ) &=& \frac{2}{\pi} \, \Gamma(n+2) \\ \nonumber
& \times& {\Im}{\mathrm{m}} \left[ i^{n+2} \lim_{\xi \to 0^+} \int_0^{\infty} dk \, \tilde{f}(k) \, \frac{k}{ (k+i \xi)^{n+2} } \right] 
\ea
which can also be expressed as
\begin{equation}
( r^n, f ) = \lim_{Q \to \infty}( r^n, f )_{_Q}
\end{equation}
with 
\ba
( r^n, f )_{_Q} &=& \frac{2}{\pi} \, \Gamma(n+2) \label{rqtrunc} \\ \nonumber
& \times &  {\Im}{\mathrm{m}} \left[ i^{n+2} \lim_{\xi \to 0+} \int_0^{Q} dk \, \tilde{f}(k) \, \frac{k}{ (k+i \xi)^{n+2} } \right] 
\ea
by inverting the $Q$ and $\xi$ limits. 
The distribution $(k+~i0)^{-n-2}$ is defined for positive real numbers as~\cite{Gue62}
\begin{equation}
\label{distrib}
(k+i0)^{-n-2} = k_+^{-n-2}  - i \, \frac{\pi}{2} \, \frac{(-)^{n+1}}{(n+1)!} \, \delta^{(n+1)}(k)
\end{equation}
where $k_+^{-n-2}$ is the usual distribution as defined in~ Ref.~\cite{Gue62}
and $\delta^{(n+1)}$ is the $(n+1)^{\mathrm{th}}$ derivative of the $\delta$-distribution. Particularly, 
\begin{equation}
\label{delta}
\hspace*{-2pt} \left( \delta^{(n+1)},k \tilde f(k) \right) = (-1)^{n+1} \, \left( \delta , \left[ k\tilde f(k) \right]^{(n+1)} \right) 
\end{equation}
which inserted in Eq.~\eqref{rqtrunc} through Eq.~\eqref{distrib} yields
\begin{equation}
( r^n, f )_{_Q} = \mathcal{A} \, {\Im}{\mathrm{m}} \left[ i^{n+2} \right] + \mathcal{B} \, {\Im}{\mathrm{m}} \left[ i^{n+1} \right] \label{TheQ}
\end{equation}
with
\begin{eqnarray}
\mathcal{A} & = & \frac{2}{\pi} \, \Gamma(n+2) \\
& \times & \left[ \int_0^{Q} dk \, \frac{ k \tilde{f}(k) - F(k)}{ k^{n+2} } - \int_Q^{\infty} dk \, \frac{F(k)}{k^{n+2}} \right] \nonumber \\
\mathcal{B} & = & \left. \left( k \tilde{f}(k) \right)^{(n+1)} \right\vert_{k=0}
\end{eqnarray}
where 
\be
F(k) = \sum_{j=0}^{(n-1)/2} g_j \, k^{2j+1}
\ee
is the Mac Laurin development of the odd function $k \tilde{f}(k)$. It is obvious from Eq.~\eqref{TheQ} that the $\mathcal{A}$-term contributes only to odd integer moments ($n$=$2p$-$1$) and that the $\mathcal{B}$-term contributes solely to even integer moments ($n$=$2p$), such that 
\begin{equation}
( r^{2p}, f )_{_Q} = (-1)^p \left. \left( k \tilde{f}(k) \right)^{(2p+1)} \right\vert_{k=0} \, .
\end{equation}
This last equation establishes that even truncated moments for $Q$$\neq$0 are independent of the truncation cut-off, and are consequently strictly equal to the infinite moment $(r^{2p}, f )$ for any form factor function $\tilde{f}(k)$.

\section{Relativistic moments} \label{momrel}

According to the prescription of Ref.~\cite{Che23}, the moments of the nucleon charge density can be expressed in the Breit frame considering the correction factor
\begin{equation}
f_{rel.} = { \left[ 1 + \frac{k^2}{4M^2} \right] }^{-1/2} 
\end{equation}
which takes into account the relativistic effects arising form the nucleon motion, and yields the substitution
\begin{equation}
\widetilde{G}_E (k^2) = f_{rel} \, G_E (k^2) \, .
\end{equation}
Particularly, the truncated moments of Eq.~\eqref{eqIMT} becomes 
\begin{equation}
\langle r^{\lambda} \rangle_{_Q} = \frac{2}{\pi}  \, \Gamma(\lambda+2) \, 
\lim_{\xi \to 0^+} \int_{0}^{Q} \mathrm{d}k \, {\mathcal I}_{\lambda}(k,\xi) \, \widetilde{G}_E (k^2) \, .
\end{equation}
We denote
\begin{eqnarray}
\tau_Q & = & Q^2 / 4 M^2 \\
\tau_{\Lambda} & = & \Lambda^2 / 4 M^2 
\end{eqnarray}
with
\begin{equation} 
\Lambda = i k_n 
\end{equation}
where $k_n$ ($n$=$1,2,3$) are the poles of the form factor function (Eq.~\eqref{ffefun}). Following the monopole decomposition technique of Sec.~\ref{section3}, Eqs.~\eqref{NegMom}-\eqref{2pmom} become
\begin{eqnarray}
& & \left( r^{-2},\rho_{M(\Lambda)} \right)_{_Q} = \frac{\Lambda^2}{2 \sqrt{1-\tau_\Lambda}} \\
& & \,\,\, \times \, \left\{ \ln{\left( \frac{1 - \sqrt{\frac{1+\tau_Q}{1-\tau_\Lambda}}}{1 + \sqrt{\frac{1+\tau_Q}{1-\tau_\Lambda}}} \right)} + \ln{\left( \frac{1 + \sqrt{\frac{1}{1-\tau_\Lambda}}}{1 - \sqrt{\frac{1}{1-\tau_\Lambda}}} \right)} \right\} \nonumber \\
& & \left( r^{2p-1},\rho_{M(\Lambda)} \right)_{_Q} = \frac{2}{\pi} \, \frac{(2p)!}{\Lambda^{2p-1}} \, \left\{  \frac{1}{\sqrt{1-\tau_\Lambda}} \right. \\
& & \,\,\, \times \, {\mathrm{Arctan}{\left( \frac{Q}{\Lambda} \sqrt{\frac{1-\tau_\Lambda}{1+\tau_Q}} \right)} } + \sqrt{1+\tau_Q} \, \sum_{j=1}^{p} \frac{(-1)^{j-1}}{2j-1}\nonumber \\
& & \,\,\, \times \, \left. {\left( \frac{\Lambda}{Q} \right)}^{2j-1} {_2F_1}\left( 1,1-j;\frac{3}{2}-j; -\tau_Q \right) \right\} \nonumber \\
& & \left( r^{2p},\rho_{M(\Lambda)} \right)_{_Q} = \frac{(2p+1)!}{(\Lambda)^{2p}} \, \left\{ \frac{1}{\sqrt{1-\tau_\Lambda}} \right. \label{MomEvRel} \\
& & \,\,\, - \, \left. \frac{(2p+1)!!}{(2p+2)!!} \, {\left( \tau_\Lambda \right)}^{p+1} \, {_2F_1}\left( 1,p+\frac{3}{2};p+2;\tau_\Lambda \right) \right\} \nonumber
\end{eqnarray}
where $_2F_1(a,b;c;z)$ is the hypergeometric function.
%
%

\bibliography{MomDenAppPRC}

%
%

\end{document}